\documentclass[10pt,epsfig]{article}  
\usepackage{epsfig}
\usepackage[body={6.5in, 9.0in},left=1.0in, top = 1.0in]{geometry}
\usepackage{amsmath}
\usepackage{amssymb}
\usepackage{graphicx}
\usepackage{appendix}
\begin{document}
\newcommand{\sheptitle}
{On the ambiguities in the tri-bimaximal mixing matrix and 
 corresponding charged lepton corrections }
\newcommand{\shepauthor}
{Chandan Duarah$^{a,b}$\footnote{Corresponding author : chandan.duarah@gmail.com} 
and N Nimai Singh$^b$}
\newcommand{\shepaddress}
   {$^a$Department of Physics, Dibrugarh University, Dibrugarh-786 004, India \\
   $^b$Department of Physics, Gauhati University,
     Guwahati-781 014, India}
\newcommand{\shepabstract}
{Two negative signs naturally appear in the $U_{\mu 1}$ and $U_{\tau 2}$
elements of the Tri-bimaximal (TBM) matrix for positive values of the
mixing angles $\theta_{12}$ and
$\theta_{23}$. Apart from this, in other TBM matrices
negative signs are shifted to other elements in each
case. They account for positive as well as negative
values of $\theta_{12}$ and $\theta_{23}$.
We discuss the sign ambiguity in the TBM matrix and find that
the TBM matrices, in fact, can be divided into two groups under
certain circumstances. Interestingly, this classification of TBM matrices
is accompanied by two different $\mu-\tau$ symmetric mass matrices which
can separately be related to the groups. To accommodate non-zero value of $\theta_{13}$
and deviate $\theta_{23}$ towards first octant, we then perturb the TBM mixing 
ansatz with the help of
charged lepton correction. The diagonalizing matrices for charged lepton 
mass matrices also possess sign ambiguity and respect the grouping
of TBM matrices. They
are parameterized in terms
of the Wolfenstein parameter $\lambda$ and satisfy unitarity condition
up to second order in $\lambda$.\\

Key-words : Tri-bimaximal mixing, Mu-tau symmetry,
           Charged lepton correction\\
PACS No. : 14.60.Pq}

\begin{titlepage}
\begin{center}
{\large{\bf\sheptitle}}
\bigskip\\
\shepauthor
\\
{\it\shepaddress}
\bigskip\\
{\bf Abstract}
\bigskip
\end{center}
\setcounter{page}{0}
\shepabstract
\end{titlepage}

\section{Introduction}

Tri-bimaximal(TBM) mixing, also known as HPS (Harrison-Perkins-Scott) mixing [1], 
is a specific lepton mixing ansatz which
draws special interest in the search of the exact lepton mixing pattern. 
It respects $\mu-\tau$ symmetry [2] and can also be realised from 
discrete symmetries like $A_4$, $S_4$ [3-6]. These interesting facts add
significant attention to TBM mixing ansatz. Except the prediction $\theta_{13}=0$
on the reactor angle, other two predictions on solar angle $\theta_{12}$ and
atmospheric angle $\theta_{23}$ of TBM mixing are attractively close to
existing global data. However a small non zero value of $\theta_{13}$, confirmed by
recent results from DAYA BAY [7], RENO [8] and DOUBLE CHOOZ [9] collaborations,
indicates certain deviation of neutrino mixing from the exact TBM mixing
ansatz. The global analysis of $3\nu$ oscillation data [10] prefers first octant
 for $\theta_{23}$.
A lot of works which discusses deviations from
TBM mixing is found in the literature [11-13].
We address the issue of sign ambiguity in the TBM mixing matrix and
suitable charged lepton correction to TBM mixing, which can accommodate
non zero $\theta_{13}$ and $\tan^2\theta_{23}<1$ as well.\\

In TBM mixing ansatz neutrino mass eigenstate $\nu_2$ is
tri-maximally mixed between all three lepton flavours while the mass eigenstate
$\nu_3$ is bimaximally mixed between $\nu_{\mu}$ and $\nu_{\tau}$ flavors. Consequences of this
mixing ansatz are: $\theta_{13}=0$, $\theta_{23}=\pm 45^{\circ}$ and 
$\theta_{12}=\pm \sin^{-1}(\frac{1}{\sqrt{3}})$.
When all the elements
of the mixing matrix are expressed in moduli squares TBM matrix has the
form [1]

\begin{equation}
\left( |U_{TB}|^2 \right)= \begin{pmatrix}
 \frac{2}{3} & \frac{1}{3} & 0 \\
 \frac{1}{6} & \frac{1}{3} & \frac{1}{2} \\
 \frac{1}{6} & \frac{1}{3} & \frac{1}{2} \\
\end{pmatrix}.
\end{equation}

Taking square root of each element of eq.(1), TBM matrix $U_{TB}$
can be obtained where each element can assume either positive or 
negative value. The choice of the sign is not unique; rather it arises
from the particular model considered. A few familiar choices, available
in the literature, for $U_{TB}$ are 

\begin{equation}
 \begin{pmatrix}
          \sqrt{\frac{2}{3}} &  \frac{1}{\sqrt{3}} & 0 \\
           -\frac{1}{\sqrt{6}}  & \frac{1}{\sqrt{3}}  & \frac{1}{\sqrt{2}} \\
          \frac{1}{\sqrt{6}}  & -\frac{1}{\sqrt{3}}   & \frac{1}{\sqrt{2}} \\

                \end{pmatrix},
\end{equation}
\begin{equation}
 \begin{pmatrix}
          \sqrt{\frac{2}{3}} &  \frac{1}{\sqrt{3}} & 0 \\
           -\frac{1}{\sqrt{6}}  & \frac{1}{\sqrt{3}}  & -\frac{1}{\sqrt{2}} \\
          -\frac{1}{\sqrt{6}}  & \frac{1}{\sqrt{3}}   & \frac{1}{\sqrt{2}} \\

                \end{pmatrix},
\end{equation}

\begin{equation}
 \begin{pmatrix}
          \sqrt{\frac{2}{3}} &  \frac{1}{\sqrt{3}} & 0 \\
           -\frac{1}{\sqrt{6}}  & \frac{1}{\sqrt{3}}  & \frac{1}{\sqrt{2}} \\
          -\frac{1}{\sqrt{6}}  & \frac{1}{\sqrt{3}}   & -\frac{1}{\sqrt{2}} \\

                \end{pmatrix},
\end{equation}
\begin{equation}
 \begin{pmatrix}
          \sqrt{\frac{2}{3}} &  -\frac{1}{\sqrt{3}} & 0 \\
           \frac{1}{\sqrt{6}}  & \frac{1}{\sqrt{3}}  & -\frac{1}{\sqrt{2}} \\
          \frac{1}{\sqrt{6}}  & \frac{1}{\sqrt{3}}   & \frac{1}{\sqrt{2}} \\

                \end{pmatrix}
\end{equation}
and
\begin{equation}
\begin{pmatrix}
          -\sqrt{\frac{2}{3}} &  \frac{1}{\sqrt{3}} & 0 \\
           \frac{1}{\sqrt{6}}  & \frac{1}{\sqrt{3}}  & \frac{1}{\sqrt{2}} \\
          \frac{1}{\sqrt{6}}  & \frac{1}{\sqrt{3}}   & -\frac{1}{\sqrt{2}} \\

                \end{pmatrix},
\end{equation}

where positions of negative signs are different in different matrices.
Starting from the standard Particle Data Group (PDG) parametrization of the
lepton mixing matrix it can be shown that above TBM matrices corresponds to
different choices of positive and negative values of the mixing angles
$\theta_{12}$ and $\theta_{23}$. The sign ambiguity in the TBM matrix
sometimes create inconveniences in phenomenological works [14],
related to parameterization
of neutrino mass matrices. Different choices of TBM matrices lead to different results.
In order to avoid such inconveniences we are
motivated to place different TBM
matrices in two groups viz. Group-I and Group-II. It is interesting to see that
 this classification is 
immediately followed by the
identification of two different $\mu-\tau$ symmetric mass matrices which
are separately associated with the groups. They differ from each other
by a distinguishing character obeyed by $m_{e\mu}$ and $m_{e\tau}$
elements of the mass matrix. Group I contains a single 
TBM matrix which accounts for positive values of both $\theta_{12}$ and 
$\theta_{23}$ while group-II contains other TBM matrices which account
for both positive and negative values of mixing angles. In regard of 
phenomenological works [14], this classification
then directs us to relate the TBM matrix of group-I, say, only with the mass 
matrix that is associated with the same group. This omits misleading results
in numerical analysis. The classification is also suitable in the discussion of
charged lepton correction to TBM mixing. We find an appropriate
form for the diagonalizing matrix of the charged lepton mass matrix
 which can generate
non zero value of $\sin \theta_{13}$ and $\tan^2 \theta_{23}<1$.
These charged lepton mass diagonalizing matrices also reflect sign
ambiguity and two different diagonalizing matrices separately work for the
two groups.\\

The paper is organised as follow: In sec.(2) we discuss the sign ambiguity
and the classification of TBM matrices. Sec.(3) presents charged lepton  
correction to TBM mixing without CP effects. The discussion of charged
lepton correction is reanalyzed in presence of a CP violating phase in
sec.(4). Finally sec.(5) is devoted to summary and discussion.\\

\section{TBM mixing matrix and the sign ambiguity}  

In the standard PDG parametrization[15], the lepton mixing
matrix, also known as PMNS (Pontecorvo-Maki-Nakagawa-Sakata) matrix is 
written as 
\begin{equation}
       U_{PMNS} = \begin{pmatrix}
    c_{12} c_{13}                       & s_{12} c_{13} 
                                                          & s_{13} e^{-i \delta_{CP}}\\
    -s_{12} c_{23}-c_{12} s_{23} s_{13}e^{i \delta_{CP}}
     & c_{12} c_{23}-s_{12} s_{23} s_{13} e^{i \delta_{CP}}
                                                          & s_{23} c_{13}\\
    s_{12} s_{23}-c_{12} c_{23} s_{13}e^{i \delta_{CP}} 
    & -c_{12} s_{23}-s_{12} c_{23} s_{13} e^{i \delta_{CP}} 
                                                          & c_{23} c_{13} \\ 
                                                      \end{pmatrix}.P ,
\end{equation}
where $c_{ij}=\cos \theta_{ij}$ ,
$s_{ij}=\sin \theta_{ij}$ with $i,j=1,2,3$, $\delta_{CP}$ 
is the Dirac CP phase and $P=diag(1, e^{i \alpha}, e^{i \beta})$
is the diagonal matrix which contains two Majorana CP phases $\alpha$ and $\beta$.
In our discussion we ignore Majorana phases. \\

For TBM mixing $s_{13}=0$ and under this condition eq.(7) reduces to
\begin{equation}
       U_{PMNS} = \begin{pmatrix}
    c_{12}                       & s_{12} 
                                                          & 0\\
    -s_{12} c_{23}
     & c_{12} c_{23}
                                                          & s_{23} \\
    s_{12} s_{23} 
    & -c_{12} s_{23} 
                                                          & c_{23}  \\ 
                                                      \end{pmatrix}.
\end{equation}

We would now like to classify different TBM matrices presented in eqs.(2-6).
The TBM matrix in eq.(2) is placed in Group-I and we
denote it as $U^{I}_{TB}$. This TBM matrix can be obtained from eq.(8) for positive values of both
$\theta_{12}$ and $\theta_{23}$. 
Remaining four TBM matrices in eqs.(3-6) are placed in group-II and we denote them as
$U^{IIa}_{TB}$, $U^{IIb}_{TB}$, $U^{IIc}_{TB}$ and $U^{IId}_{TB}$ respectively.
The TBM matrix $U^{IIa}_{TB}$ in eq.(3) can be obtained from eq.(8) for  
positive $\theta_{12}$
and negative $\theta_{23}$ and  $U^{IIc}_{TB}$ in eq.(5)  can be obtained 
from eq.(8) for negative values of both $\theta_{12}$
and  $\theta_{23}$. TBM matrix  $U^{IId}_{TB}$ in eq.(6) 
can be obtained from eq.(8) 
under the transformations 
  $\theta_{12}\longrightarrow (\pi -\theta_{12})$ and
  $\theta_{23}\longrightarrow (\pi -\theta_{23})$.\\
  
For convenience let us now represent these choices of TBM matrices as different 
sign conventions where, for example, we get the convention
$$ \begin{pmatrix}
          + &  + & 0 \\
           -  & +  & + \\
          +  & -   & + \\

                \end{pmatrix}, $$

for the TBM matrix in eq.(2). We would like to use a bold face notation 
${\bf U^I_{TB}}$, in correspondence to $ U^I_{TB}$, for this sign convention. Similarly we get sign conventions
${\bf U^{IIa}_{TB}}$, ${\bf U^{IIb}_{TB}}$, ${\bf U^{IIc}_{TB}}$ and ${\bf U^{IId}_{TB}}$
for the TBM matrices in eqs.(3-6) respectively.


The underlying motivation for this classification of TBM matrices is
basically extracted from phenomenological works [14], based on 
parameterization of $\mu-\tau$ symmetric mass matrices. The general
 $\mu-\tau$ symmetric mass matrix  is given by
\begin{equation}
m^{II}_{\mu \tau}=\begin{pmatrix}
A & B & B \\
B & C & D \\
B & D & C \\
\end{pmatrix},
\end{equation}
which leads to maximal
atmospheric mixing and zero reactor angle, leaving behind the solar
angle arbitrary. Choosing the diagonalizing matrix 
\begin{equation}
U = \begin{pmatrix}
          c_{12} &  s_{12} & 0 \\
           -\frac{s_{12}}{\sqrt{2}}  & \frac{c_{12}}{\sqrt{2}}  & -\frac{1}{\sqrt{2}} \\
           -\frac{s_{12}}{\sqrt{2}}  & \frac{c_{12}}{\sqrt{2}}   & \frac{1}{\sqrt{2}} \\

                \end{pmatrix}
\end{equation}
for the mass matrix in eq.(9), we get the important relation
\begin{equation}
\tan 2 \theta_{12}= \vert \frac{2\sqrt{2}B}{A-C-D} \vert,
\end{equation}

which allows us to fix the value of solar angle at any desired value by the choice
of the elements of the mass matrix. This relation has significant implications
in works done in ref.[14].
Here we have followed the diagonalizaton
relation $m^{II diag}_{\mu \tau}=U^T m^{II}_{\mu \tau} U$ and for the diagonalizing
matrix in eq.(10) we follow the sign convention ${\bf U^{IIa}_{TB}}$.
The inconvenience due to sign ambiguity in TBM matrix is that if we choose the
sign convention ${\bf U^{I}_{TB}}$, for example, for the diagonalizing matrix
 $U$ to diagonalize
the mass matrix in eq.(9), we get undesired relation for $\tan 2 \theta_{12}$
instead of eq.(11).\\

We note that $ U^I_{TB}$ in Group I predicts both the mixing angles $\theta_{12}$ and 
$\theta_{23}$ positive while the TBM matrices in Group II predict either one of the 
mixing angles negative or both negative. The $\mu-\tau$ symmetric mass matrix
which is consistent with positive mixing angles is given by [2]

\begin{equation}
m^{I}_{\mu \tau}=\begin{pmatrix}
A & -B & B \\
-B & C & D \\
B & D & C \\
\end{pmatrix}.
\end{equation}

This mass matrix is different from $m^{II}_{\mu \tau}$ by the distinguishing
character $m_{e \mu}=-m_{e \tau}$.
We therefore associate this mass matrix with the TBM matrix of Group I. 
Then the diagonalizing matrix in eq.(10),
along with sign convention ${\bf U^I_{TB}}$, leads to the desired expression 

\begin{equation}
\tan 2 \theta_{12}= \vert \frac{2\sqrt{2}B}{A-C+D} \vert.
\end{equation}

The mass matrix $m^{II}_{\mu \tau}$ does not gaurantee positive mixing angles
and it works for the TBM matrices of Group II. We obtain the same expression
for $\tan 2 \theta_{12}$, given in eq.(11),
when we follow any sign convention ${\bf U^{IIi}_{TB}}$ ($i=a,b,c,d$) for the 
diagonalizing matrix $U$ in eq.(10).\\

\begin{table}
\begin{center}
\caption{Best fit, $1\sigma$ and $3\sigma$ ranges of parameters for NH obtained
         from global analysis by Forero {\it et al.}[10]}
\begin{tabular}{ccccccc}
\hline
 & best fit  & $1\sigma$ range & $3\sigma$ range \\  
  \hline                                                 
$\tan^2 \theta_{12}$ &  0.470 &  0.435-0.506 & 0.370-0.587  \\
$\tan^2 \theta_{23}$ & 0.745 &  0.667-0.855 & 0.563-2.125 \\
$\sin^2 \theta_{13}$ & 0.0246 & 0.0218-0.0275 & 0.017-0.033 \\
\hline
\end{tabular}
\end{center}
\end{table}

\section{Charged lepton correction to TBM matrix} 


Charged lepton corrections [16-19] to neutrino mixing may be defined through the
relation 
\begin{equation}
U_{PMNS}= U_{lL}^{\dagger} U_{\nu},
\end{equation} 
where  $U_{PMNS}$ is the lepton mixing matrix,
$U_{lL}$ and $U_{\nu}$ are the diagonalizing matrices 
for charged lepton and left-handed Majorana neutrino mass matrices
respectively. They are defined through the relations :
$m_l=U_{lL} m_l^{diag} V_{lR}^{\dagger}$ 
and $m_{\nu}=U_{\nu}^{*}m_{\nu}^{diag} U_{\nu}^{\dagger}$,
where $m_l^{diag}=Diag(m_e,m_{\mu},m_{\tau})$
and $m_{\nu}^{diag}=Diag(m_1,m_2,m_3)$. In the basis where charged lepton mass matrix
$m_l$ is diagonal, 
 $U_{PMNS}=U_{\nu}$, $U_{lL}$ being identity matrix. Effects of charged lepton correction
in this basis can be absorbed in
the left-handed Majorana mass matrix as
$m_{\nu}^{\prime}=U_{lL}^{\dagger} m_{\nu} U_{lL}$. \\

For our case $U_{\nu}$ is to be given by $U_{TB}$. We then propose a
possible form for the charged lepton mass diagonalizing matrix $U_{lL}$,
parameterized in terms of Wolfenstein parameter $\lambda$ [20], which can
generate non zero $\theta_{13}$ as well as $\tan^2 \theta_{23}<1$.
In our analysis we are preferring first octant for $\theta_{23}$,
motivated by the global analysis data [10]. The diagonalizing matrix is given by

\begin{equation}
U^I_{lL}=\begin{pmatrix}
1-\frac{\lambda^2}{4} & \frac{\lambda}{2} & \frac{\lambda}{2} \\
-\frac{\lambda}{2} & 1-\frac{\lambda^2}{8} & \frac{7}{8}\lambda^2 \\
-\frac{\lambda}{2} & -\frac{9}{8}\lambda^2 & 1-\frac{\lambda^2}{8} \\
\end{pmatrix},
\end{equation}
which works for the TBM matrix of Group-I. The diagonalizing
matrix that works for TBM matrices of Group-II is given by 
\begin{equation}
U^{II}_{lL}=\begin{pmatrix}
1-\frac{\lambda^2}{4} & \frac{\lambda}{2} & -\frac{\lambda}{2} \\
-\frac{\lambda}{2} & 1-\frac{\lambda^2}{8} & -\frac{7}{8}\lambda^2 \\
\frac{\lambda}{2} & \frac{9}{8}\lambda^2 & 1-\frac{\lambda^2}{8} \\
\end{pmatrix}.
\end{equation}

The diagonalizing matrices in eqs.(15) and (16)
satisfy unitarity condition up to second order in $\lambda$. Their structures can be
derived from the diagonalizing matrix considered in Ref.[19], which is

\begin{equation}
U^{\dagger}_{lL}= \tilde{R}^{\dagger}_{23} \tilde{U}^{\dagger}_{lL}.
\end{equation}
Here 
\begin{equation}
\tilde{R}_{23}= \begin{pmatrix}
           1  &  0   &  0 \\
           0  & \tilde{c}_{23} & \tilde{s}_{23}\\
           0  & -\tilde{s}_{23} & \tilde{c}_{23}\\
\end{pmatrix}
\end{equation}
and
\begin{equation}
\tilde{U}_{lL}=\begin{pmatrix}
1-\frac{\lambda^2}{4} & -\frac{\lambda}{2} & -\frac{\lambda}{2} \\
 \frac{\lambda}{2} & 1-\frac{\lambda^2}{8} & -\frac{\lambda^2}{8} \\
 \frac{\lambda}{2} & -\frac{\lambda^2}{8} & 1-\frac{\lambda^2}{8} \\
 \end{pmatrix}.
\end{equation}

Eqs.(17), (18) and (19) then gives
\begin{equation}
U^{\dagger}_{lL}=\begin{pmatrix}
1-\frac{\lambda^2}{4} & \frac{\lambda}{2} & \frac{\lambda}{2} \\
-(\tilde{c}_{23}-\tilde{s}_{23})\frac{\lambda}{2} 
  & \tilde{s}_{23}\frac{\lambda^2}{8} + \tilde{c}_{23}(1-\frac{\lambda^2}{8})
  & -\tilde{c}_{23}\frac{\lambda^2}{8} - \tilde{s}_{23}(1-\frac{\lambda^2}{8})\\
  -(\tilde{c}_{23}+\tilde{s}_{23})\frac{\lambda}{2}
  & -\tilde{c}_{23}\frac{\lambda^2}{8} + \tilde{s}_{23}(1-\frac{\lambda^2}{8})
  & -\tilde{s}_{23}\frac{\lambda^2}{8} + \tilde{c}_{23}(1-\frac{\lambda^2}{8}) \\
\end{pmatrix}.
\end{equation}
Under the approximations $\tilde{s}_{23}\approx \lambda^2$ and $\tilde{c}_{23}\approx 1$
eq.(20) leads to
\begin{equation}
U^{\dagger}_{lL} \approx \begin{pmatrix}
1-\frac{\lambda^2}{4} & \frac{\lambda}{2} & \frac{\lambda}{2} \\
-(1-\lambda^2)\frac{\lambda}{2} & \frac{\lambda^4}{8} + (1-\frac{\lambda^2}{8})
                & -\frac{\lambda^2}{8} - \lambda^2(1-\frac{\lambda^2}{8}) \\
  -(1+\lambda^2)\frac{\lambda}{2}  & -\frac{\lambda^2}{8} + \lambda^2(1-\frac{\lambda^2}{8}) 
          & -\frac{\lambda^4}{8} + (1-\frac{\lambda^2}{8}) \\
\end{pmatrix}.
\end{equation}

Or
\begin{equation}
U^{\dagger}_{lL} = \begin{pmatrix}
1-\frac{\lambda^2}{4} & \frac{\lambda}{2} & \frac{\lambda}{2} \\
-\frac{\lambda}{2} & 1-\frac{\lambda^2}{8} &  -\frac{9}{8}\lambda^2  \\
-\frac{\lambda}{2} & \frac{7}{8}\lambda^2 & 1-\frac{\lambda^2}{8} \\
\end{pmatrix} + O(\lambda^{n>2}).
\end{equation}
The structure of the matrix on the right hand side of eq.(22) is what was considered
for $U^I_{lL}$ or $U^{II}_{lL}$. \\
 
 Then the relations   
$U^I_{PMNS}= (U_{lL}^I)^{\dagger} U^I_{TB}$ and
$U^{II}_{PMNS}= (U_{lL}^{II})^{\dagger} U^{IIi}_{TB}$ ($i=a,b,c,d$) lead to the PMNS matrices

\begin{equation}
U^I_{PMNS}=\begin{pmatrix}
          \sqrt{\frac{2}{3}}(1-\frac{\lambda^2}{4}) 
          &  \frac{1}{\sqrt{3}}(1-\frac{\lambda^2}{4}) 
          & -\frac{\lambda}{\sqrt{2}} \\
           -\frac{1}{\sqrt{6}}(1-\lambda+\lambda^2) 
            & \frac{1}{\sqrt{3}}(1+\frac{\lambda}{2}+\lambda^2) 
             & \frac{1}{\sqrt{2}}(1-\frac{5}{4}\lambda^2) \\
          \frac{1}{\sqrt{6}}(1+\lambda-\lambda^2) 
           & -\frac{1}{\sqrt{3}}(1-\frac{\lambda}{2}-\lambda^2)  
            & \frac{1}{\sqrt{2}}(1+\frac{3}{4}\lambda^2) \\              
             \end{pmatrix},
\end{equation}

\begin{equation} 
U^{IIa}_{PMNS}=\begin{pmatrix}
          \sqrt{\frac{2}{3}}(1-\frac{\lambda^2}{4}) 
          &  \frac{1}{\sqrt{3}}(1-\frac{\lambda^2}{4}) 
          & \frac{\lambda}{\sqrt{2}} \\
           -\frac{1}{\sqrt{6}}(1-\lambda+\lambda^2) 
            & \frac{1}{\sqrt{3}}(1+\frac{\lambda}{2}+\lambda^2) 
             & -\frac{1}{\sqrt{2}}(1-\frac{5}{4}\lambda^2) \\
          -\frac{1}{\sqrt{6}}(1+\lambda-\lambda^2) 
           & \frac{1}{\sqrt{3}}(1-\frac{\lambda}{2}-\lambda^2)  
            & \frac{1}{\sqrt{2}}(1+\frac{3}{4}\lambda^2) \\              
             \end{pmatrix},
\end{equation}
\begin{equation} 
U^{IIb}_{PMNS}=\begin{pmatrix}
          \sqrt{\frac{2}{3}}(1-\frac{\lambda^2}{4}) 
          &  \frac{1}{\sqrt{3}}(1-\frac{\lambda^2}{4}) 
          & -\frac{\lambda}{\sqrt{2}} \\
           -\frac{1}{\sqrt{6}}(1-\lambda+\lambda^2) 
            & \frac{1}{\sqrt{3}}(1+\frac{\lambda}{2}+\lambda^2) 
             & \frac{1}{\sqrt{2}}(1-\frac{5}{4}\lambda^2) \\
          -\frac{1}{\sqrt{6}}(1+\lambda-\lambda^2) 
           & \frac{1}{\sqrt{3}}(1-\frac{\lambda}{2}-\lambda^2)  
            & -\frac{1}{\sqrt{2}}(1+\frac{3}{4}\lambda^2) \\              
             \end{pmatrix},
\end{equation}
\begin{equation} 
U^{IIc}_{PMNS}=\begin{pmatrix}
          \sqrt{\frac{2}{3}}(1-\frac{\lambda^2}{4}) 
          &  -\frac{1}{\sqrt{3}}(1-\frac{\lambda^2}{4}) 
          & \frac{\lambda}{\sqrt{2}} \\
           \frac{1}{\sqrt{6}}(1+\lambda+\lambda^2) 
            & \frac{1}{\sqrt{3}}(1-\frac{\lambda}{2}+\lambda^2) 
             & -\frac{1}{\sqrt{2}}(1-\frac{5}{4}\lambda^2) \\
          \frac{1}{\sqrt{6}}(1-\lambda-\lambda^2) 
           & \frac{1}{\sqrt{3}}(1+\frac{\lambda}{2}-\lambda^2)  
            & \frac{1}{\sqrt{2}}(1+\frac{3}{4}\lambda^2) \\              
             \end{pmatrix},
\end{equation}
and
\begin{equation} 
U^{IId}_{PMNS}=\begin{pmatrix}
          -\sqrt{\frac{2}{3}}(1-\frac{\lambda^2}{4}) 
          &  \frac{1}{\sqrt{3}}(1-\frac{\lambda^2}{4}) 
          & -\frac{\lambda}{\sqrt{2}} \\
           \frac{1}{\sqrt{6}}(1-\lambda+\lambda^2) 
            & \frac{1}{\sqrt{3}}(1+\frac{\lambda}{2}+\lambda^2) 
             & \frac{1}{\sqrt{2}}(1-\frac{5}{4}\lambda^2) \\
          \frac{1}{\sqrt{6}}(1+\lambda-\lambda^2) 
           & \frac{1}{\sqrt{3}}(1-\frac{\lambda}{2}-\lambda^2)  
            & -\frac{1}{\sqrt{2}}(1+\frac{3}{4}\lambda^2) \\              
             \end{pmatrix}
\end{equation}

respectively. All these matrices predict

\begin{equation}
\sin^2 \theta_{13} = \displaystyle \vert \frac{\lambda}{\sqrt{2}}\vert^2,
\end{equation}
\begin{equation}
\tan^2 \theta_{12} =0.5,
\end{equation}
\begin{equation}
\tan^2 \theta_{23}=\displaystyle \vert 
 \frac{(1-\frac{5}{4}\lambda^2)}{(1+\frac{3}{4}\lambda^2)}\vert^2 .
\end{equation}

For $\lambda=0.225$ we get $\sin^2 \theta_{13}\approx 0.025$ and 
$\tan^2 \theta_{23}\approx 0.81$. These predictions on $\sin^2 \theta_{13}$
and $\tan^2 \theta_{23}$ are consistent with $1\sigma$ range of global data (Table 1).\\

The diagonalizing matrices in eqs.(15) and (16) also possess sign ambiguity
and their identification for the two groups 
is analogous to the case of $\mu-\tau$ symmetric mass matrices. We want to 
emphasize that if we follow
the relation $U^I_{PMNS}= (U_{lL}^I)^{\dagger} U^{IIi}_{TB}$
instead of $U^I_{PMNS}= (U_{lL}^I)^{\dagger} U^I_{TB}$ say, 
it alters all the predictions presented in eqs.(28-30).\\

It is important to note here that the PMNS matrix in any of the eqs.(23-27)
when compared with the Tri-bimaximal-Cabibbo mixing matrix ($U_{TBC}$)
proposed by King [21], we find that it can predict $\tan^2 \theta_{23}<1$
along with non zero $\theta_{13}$ while $U_{TBC}$ predicts non zero $\theta_{13}$
keeping the solar and atmospheric angles fixed at TBM values.

\section{CP violation}

The PMNS mixing matrices in eqs.(23-27) conserves CP symmetry. In this
section we would like to analyze the effects of a CP violating phase $\delta$
on the predictions of the PMNS matrix after charged lepton correction. 
To introduce the phase $\delta$ in 
$U^{I}_{PMNS}$ we follow the Tri-bimaximal-Cabibbo mixing matrix $U_{TBC}$ [21],
given by

 \begin{equation} 
U_{TBC}=\begin{pmatrix}
          \sqrt{\frac{2}{3}}(1-\frac{\lambda^2}{4}) 
          &  \frac{1}{\sqrt{3}}(1-\frac{\lambda^2}{4}) 
          & -\frac{\lambda}{\sqrt{2}}e^{-i \delta} \\
           -\frac{1}{\sqrt{6}}(1-\lambda e^{i \delta} ) 
            & \frac{1}{\sqrt{3}}(1+\frac{\lambda}{2} e^{i \delta} ) 
             & \frac{1}{\sqrt{2}}(1-\frac{\lambda^2}{4}) \\
          \frac{1}{\sqrt{6}}(1+\lambda e^{i \delta}) 
           & -\frac{1}{\sqrt{3}}(1-\frac{\lambda}{2} e^{i \delta})  
            & \frac{1}{\sqrt{2}}(1-\frac{\lambda^2}{4}) \\              
             \end{pmatrix}
\end{equation}

If we ignore the phase $\delta$ in eq.(31) we get $U_{TBC}=\tilde{U}^{\dagger}_{lL}U_{TB}$,
where $\tilde{U}_{lL}$ is defined in eq.(19). Then from eq.(17) we find that the expression
$U_{PMNS}=U^{\dagger}_{lL}U_{TB}$ is equivalent to $U_{PMNS}=\tilde{R}^{\dagger}_{23}U_{TBC}$,
where $\tilde{R}_{23}$ is given by eq.(18). To incorporate the phase $\delta$
in PMNS matrix we therefore employ the relation
 $U_{PMNS}=\tilde{R}^{\dagger}_{23}U_{TBC}$ such that $U_{TBC}$ is now given by eq.(31)
and the approximations $\tilde{s}_{23}\approx \lambda^2$ and
 $\tilde{c}_{23}\approx 1$ should be considered in addition. 
$U_{TBC}$ in eq.(31) follows the sign convention ${\bf U^I_{TB}}$.
We thus obtain
\begin{equation}
U^{I}_{PMNS}=\begin{pmatrix}
          \sqrt{\frac{2}{3}}(1-\frac{\lambda^2}{4}) 
          &  \frac{1}{\sqrt{3}}(1-\frac{\lambda^2}{4}) 
          & -\frac{\lambda}{\sqrt{2}}e^{-i \delta} \\
           -\frac{1}{\sqrt{6}}(1-\lambda e^{i \delta}+\lambda^2) 
            & \frac{1}{\sqrt{3}}(1+\frac{\lambda}{2} e^{i \delta}+\lambda^2) 
             & \frac{1}{\sqrt{2}}(1-\frac{5}{4}\lambda^2) \\
          \frac{1}{\sqrt{6}}(1+\lambda e^{i \delta}-\lambda^2) 
           & -\frac{1}{\sqrt{3}}(1-\frac{\lambda}{2} e^{i \delta}-\lambda^2)  
            & \frac{1}{\sqrt{2}}(1+\frac{3}{4}\lambda^2) \\              
             \end{pmatrix}.
\end{equation} 

In a similar manner we obtain the PMNS matrices
for Group II as

\begin{equation} 
U^{IIa}_{PMNS}=\begin{pmatrix}
          \sqrt{\frac{2}{3}}(1-\frac{\lambda^2}{4}) 
          &  \frac{1}{\sqrt{3}}(1-\frac{\lambda^2}{4}) 
          & \frac{\lambda}{\sqrt{2}}e^{-i \delta} \\
           -\frac{1}{\sqrt{6}}(1-\lambda e^{i \delta}+\lambda^2) 
            & \frac{1}{\sqrt{3}}(1+\frac{\lambda}{2} e^{i \delta}+\lambda^2) 
             & -\frac{1}{\sqrt{2}}(1-\frac{5}{4}\lambda^2) \\
          -\frac{1}{\sqrt{6}}(1+\lambda e^{i \delta}-\lambda^2) 
           & \frac{1}{\sqrt{3}}(1-\frac{\lambda}{2}e^{i \delta}-\lambda^2)  
            & \frac{1}{\sqrt{2}}(1+\frac{3}{4}\lambda^2) \\              
             \end{pmatrix},
\end{equation}
\begin{equation} 
U^{IIb}_{PMNS}=\begin{pmatrix}
          \sqrt{\frac{2}{3}}(1-\frac{\lambda^2}{4}) 
          &  \frac{1}{\sqrt{3}}(1-\frac{\lambda^2}{4}) 
          & -\frac{\lambda}{\sqrt{2}}e^{-i \delta} \\
           -\frac{1}{\sqrt{6}}(1-\lambda e^{i \delta}+\lambda^2) 
            & \frac{1}{\sqrt{3}}(1+\frac{\lambda}{2} e^{i \delta}+\lambda^2) 
             & \frac{1}{\sqrt{2}}(1-\frac{5}{4}\lambda^2) \\
          -\frac{1}{\sqrt{6}}(1+\lambda e^{i \delta}-\lambda^2) 
           & \frac{1}{\sqrt{3}}(1-\frac{\lambda}{2} e^{i \delta}-\lambda^2)  
            & -\frac{1}{\sqrt{2}}(1+\frac{3}{4}\lambda^2) \\              
             \end{pmatrix},
\end{equation}
\begin{equation} 
U^{IIc}_{PMNS}=\begin{pmatrix}
          \sqrt{\frac{2}{3}}(1-\frac{\lambda^2}{4}) 
          &  -\frac{1}{\sqrt{3}}(1-\frac{\lambda^2}{4}) 
          & \frac{\lambda}{\sqrt{2}}e^{-i \delta} \\
           \frac{1}{\sqrt{6}}(1+\lambda e^{i \delta}+\lambda^2) 
            & \frac{1}{\sqrt{3}}(1-\frac{\lambda}{2} e^{i \delta}+\lambda^2) 
             & -\frac{1}{\sqrt{2}}(1-\frac{5}{4}\lambda^2) \\
          \frac{1}{\sqrt{6}}(1-\lambda e^{i \delta}-\lambda^2) 
           & \frac{1}{\sqrt{3}}(1+\frac{\lambda}{2} e^{i \delta}-\lambda^2)  
            & \frac{1}{\sqrt{2}}(1+\frac{3}{4}\lambda^2) \\              
             \end{pmatrix},
\end{equation}
\begin{equation} 
U^{IId}_{PMNS}=\begin{pmatrix}
          -\sqrt{\frac{2}{3}}(1-\frac{\lambda^2}{4}) 
          &  \frac{1}{\sqrt{3}}(1-\frac{\lambda^2}{4}) 
          & -\frac{\lambda}{\sqrt{2}}e^{-i \delta} \\
           \frac{1}{\sqrt{6}}(1-\lambda e^{i \delta}+\lambda^2) 
            & \frac{1}{\sqrt{3}}(1+\frac{\lambda}{2}e^{i \delta}+\lambda^2) 
             & \frac{1}{\sqrt{2}}(1-\frac{5}{4}\lambda^2) \\
          \frac{1}{\sqrt{6}}(1+\lambda e^{i \delta}-\lambda^2) 
           & \frac{1}{\sqrt{3}}(1-\frac{\lambda}{2} e^{i \delta}-\lambda^2)  
            & -\frac{1}{\sqrt{2}}(1+\frac{3}{4}\lambda^2) \\              
             \end{pmatrix}.
\end{equation}

All the PMNS matrices in eqs.(32-36) yield the same predictions of 
mixing angles as given in eqs.(28-30). Further they all
 lead to a similar expression for the rephasing
invariant quantity, defined as $J_{CP}=Im\{U_{e2}U_{\mu 3}
U^{*}_{e3}U^{*}_{\mu 2}\}$, which is 

\begin{equation}
|J_{CP}|=\displaystyle \frac{1}{6}\lambda(1+\lambda^2)(1-\frac{\lambda^2}{4})
                 (1-\frac{5}{4}\lambda^2) \sin \delta .
\end{equation}

For maximal CP violation we get $|J_{CP}|\approx 0.0364$.

\section{Summary and Discussion}

We discuss sign ambiguities in  the TBM mixing matrix which arise due
to different choices of positive and negative values of the mixing
angles $\theta_{12}$ and $\theta_{23}$. Such sign ambiguities
sometime create inconveniences in phenomenological works and numerical
analysis. To avoid the inconveniences we find it 
useful to divide different TBM matrices into two groups. Group-I
contains a single TBM matrix which accounts positive values of
both the mixing angles. Other TBM matrices are placed in Group-II.
Few of them account for positive as well as negative values of
$\theta_{12}$ and $\theta_{23}$. Some others are  found to obey
certain quadrant transformations. This grouping of TBM matrices
is followed by two $\mu-\tau$ symmetric mass matrices, separately
associated with the groups. They differ by the fact that for the
mass matrix associated with Group-I we have $m_{e\mu}=-m_{e\tau}$
while for the other, associated with Group-II, we have $m_{e\mu}=m_{e\tau}$.
The classification is also useful in the discussion of charged
lepton correction to TBM mixing. We find a possible form of the
charged lepton mass diagonalizing matrix $U_{lL}$ which can generate non zero
$\theta_{13}$ and $\tan^2 \theta_{23}<1$ consistent with latest
global analysis data. We can identify two diagonalizing
matrices, which also reflect sign ambiguities,
for the two groups of TBM matrices such that they separately
work to get desired results. The discussion of sign ambiguities
and related classifications may help authors in 
systematic phenomenological analysis. This work points out that
it is useful to do phenomenological studies related to TBM mixing
ansatz under two groups where the TBM matrix which predicts positive
mixing angles can be isolated from other TBM matrices.



\end{document}